\documentclass[aip,reprint,floatfix]{revtex4-1}

\usepackage{amsmath}
\usepackage{bm}
\usepackage{amssymb}
\usepackage{graphicx}
\usepackage{float}
\usepackage{array}
\usepackage{xcolor}

\draft 

\begin{document}



\title{Explicit screening full band quantum transport model for semiconductor nanodevices} 



\author{Yuanchen Chu}
\email[Email: ]{chu72@purdue.edu}
\author{Prasad Sarangapani}
\author{James Charles}
\affiliation{
School of Electrical and Computer Engineering, Purdue University, West Lafayette, IN 47907, USA
}
\affiliation{
Network for Computational Nanotechnology, Purdue University, West Lafayette, IN 47907, USA
}
\author{Gerhard Klimeck}
\author{Tillmann Kubis}
\affiliation{
School of Electrical and Computer Engineering, Purdue University, West Lafayette, IN 47907, USA
}
\affiliation{
Network for Computational Nanotechnology, Purdue University, West Lafayette, IN 47907, USA
}
\affiliation{
Center for Predictive Materials and Devices, Purdue University, West Lafayette, IN 47907, USA
}


\date{\today}

\begin{abstract}
State of the art quantum transport models for semiconductor nanodevices attribute negative (positive) unit charges to states of the conduction (valence) band. 
Hybrid states that enable band-to-band tunneling are subject to interpolation that yield model dependent charge contributions. 
In any nanodevice structure, these models rely on device and physics specific input for the dielectric constants.
This paper exemplifies the large variability of different charge interpretation models when applied to ultrathin body transistor performance predictions. 
To solve this modeling challenge, an electron-only band structure model is extended to atomistic quantum transport. 
Performance predictions of MOSFETs and tunneling FETs confirm the generality of the new model and its independence of additional screening models. 

\end{abstract}
\pacs{}
\maketitle 
\section{\label{sec:level1}Introduction}
As the scaling of Metal-Oxide-Semiconductor Field-Effect Transistor (MOSFET) has reached sub-10 nm regime, power consumption has become a major concern\cite{1250885}. 
The advantages of lowering the dynamic power consumption by reducing the supply voltage are fast disappearing as the static power has begun to dominate due to the exponential increase of the subthreshold leakage current\cite{Kao:2002:SLM:774572.774593}. 
Band-to-band tunneling field-effect transistor (TFET) is among the most promising candidates for future integrated circuits (ICs) due to its ability to beat the 60mV/decade limit of the subthreshold swing (SS)\cite{PhysRevLett.93.196805,1546317}.
Having a smaller subthreshold swing enables a reduction of both the supply voltage and the subthreshold leakage current, thus further lowering the power consumption of the ICs\cite{Ionescu2011}.

However, despite many predictions of outstanding device performance, most experimental TFETs underperform conventional MOSFETs\cite{6820751}. 
The discrepancy between TFET simulations and experiments indicates missing or mistreated physics in TFET simulations\cite{Mookerjea2010,7409758,7409828}.
Analytical models of TFETs indicate that the tunneling current has an exponential dependence on effective mass, energy band gap and electric field at the tunnel junction\cite{Seabaugh2010}.
Obviously, the accuracy of these quantities is crucial for quantitative prediction of TFET current-voltage characteristics. 
This is particularly true for the electrostatic landscape, since it rules all bandstructure and tunneling properties.

The Nonequilibrium Green's function (NEGF) formalism\cite{Datta2000,1224493,7605202} is widely accepted as one of the most consistent models for electronic properties in nanodevices in the presence of quantum phenomena including quantum confinement, tunneling, interferences, etc\cite{KNOCH2007572,Matyas2010}. 
Nanotransistor properties such as charge distribution and current density are commonly deduced from the NEGF equations once they are self-consistently solved with the Poisson equation.

The standard model (termed as excess-charge approach or ECA) to interpret the particle density in quantum transport calculations distinguishes the charge carrier type:
An electron (hole) in conduction (valence) band state of n-type (p-type) MOSFET is considered to contribute a negative (positive) unit charge.
This concept limits the computational load to solving electrons (holes) in the conduction (valence) band only, i.e. a few $k_BT$ of energy in addition to the energy range spanned by the applied bias voltage.
In the Poisson equation, the dielectric constant is then typically set to the material's constant. 
In tunneling devices, a particle with energy above (below) the conduction (valence) band is still considered to contribute a negative (positive) unit charge. 
For energies between the conduction and the valence band, various charge interpolation schemes exist.
As will be shown in detail in this paper, the standard model fails in various ways, both for conventional as well as for band-to-band tunneling transistors. 
It turns out - consistent with previous findings in literature\cite{PhysRevB.77.115355,Markov2015} - that the electrostatic screening of valence band electrons that do not take part in transport is device physics (and not only material) dependent.
The charge interpolation schemes required for band-to-band tunneling devices host an arbitrariness that severely limits the reliability of device performance predictions.
Any such interpolation also suffers from incompatibility with the NEGF method as discussed in detail in Appendix A. 
This is consistent with previous findings in broken-gap optoelectronic bandstructure calculations\cite{Andlauer2009}.
Therefore, the main focus of this work is the introduction of a consistent interpretation of quantum transport solutions for charge self-consistent nanodevice calculations. 

In Sec. II, the state of the art model (ECA) and a new full-band approach (FBA) for the charge self-consistent prediction of band-to-band tunneling devices are explained.
In Sec. III, results of the two models are compared. Both models agree well for situations in which the ECA model does not suffer from arbitrariness. 
The FBA model consistently describes devices beyond that application space. 
The paper is summarized in Sec. IV. 
\section{\label{sec:level1}METHOD}
All models and results of this paper are part of the multipurpose nanodevice simulation tool NEMO5\cite{6069914}.
The electronic structure is represented in atomistic nearest-neighbor tight-binding (TB) sp3d5s* basis. 
Throughout the paper, empirical TB parameters for Si, GaSb and InAs are taken from Refs.~[\onlinecite{PhysRevB.69.115201}, \onlinecite{PhysRevB.57.6493}, \onlinecite{PhysRevB.66.125207}], respectively. 
In all ultra-thin body (UTB) devices of this work, the electron transport direction ($x$) and the confinement direction ($z$) are atomically discretized.
The periodic direction ($y$) is discretized with a single unit cell and boundary conditions are applied on it.
A few examples are considered with periodicity in both, $y$- and $z$- direction. 
The full Brillouin zone of the electronic momentum ($k$) in any periodic direction is discretized.
Coherent transport is assumed and no incoherent scattering is considered.
If not explicitly stated otherwise, all results are obtained by iterating the quantum transport equations with the nonlinear Poisson equation (discretized on a three dimensional finite element grid) until convergence is achieved.
For all devices, flat band Neumann boundary conditions are assumed for the Poisson equation except for the gate/oxide interface regions where Dirichlet conditions are used to set the applied gate voltages.
Full ionization of donor and acceptor atoms is assumed. 

In the state of the art model (ECA), states that are in the conduction (valence) band for all device positions are considered to be electrons (holes) and to contribute a negative (positive) unit charge.
Since this is commonly the case for nMOSFETs or pMOSFETs, these devices allow limiting the calculations to conduction or valence bands, respectively.
In band-to-band tunneling (BTBT) situations, both valence band and conduction band states have to be considered, since states exist that overlap with conduction and valence band simultaneously. 
The density of such states is translated with a heuristic interpolation factor $\lambda$ into their charge density contribution. 
The expression for the charge density contribution of each individual lead is given by
\begin{eqnarray}
&&- q{n_{ECA}}(E,k) = qp(E,k) - qn(E,k)\nonumber\\
&&=q\left[ {1 - f(E,\mu )} \right]\lambda (E,k){\left| {\psi (E,k)} \right|^2} \nonumber \\
&&\quad -qf(E,\mu )\left[ {1 - \lambda (E,k)} \right]{\left| {\psi (E,k)} \right|^2},
\label{eq:1}
\end{eqnarray}
where $E$ is the state energy, $\psi(E,k)$ is the injected conduction or valence band state, $\mu$ is the lead Fermi level and $q$ is the positive unit charge. 
Note that in Eq.~(\ref{eq:1}) as in all subsequent equations, the position coordinates are omitted for better readability.
The electronic states $\psi(E,k)$ are solved with the quantum transmitting boundary method (QTBM)\cite{Lent1994,Luisier2006}. 
The factor $\lambda$ heuristically interpolates between the positive and negative charge interpretations of valence and conduction band. 
Therefore, $\lambda$ depends on the electrostatic potential, the conduction and valence band edges and the chosen heuristic interpolation model. 
In general $\lambda$ is a function of energy, in-plane momentum and position.
In this paper, three commonly used heuristic models for $\lambda$ have been chosen as representatives (summarized in Table.~\ref{tab:heuristics_table}).
All three heuristic models distinguish the interpolation factor $\lambda$ for energies $E$ below and above a delimiter energy $D$ given by
\begin{equation}
D = \left( {1 - \alpha } \right){E_V} + \alpha {E_C},
\label{eq:2}
\end{equation}
where $\alpha$ is a unitless number and $E_V$ ($E_C$) is the valence (conduction) band edge.
For simplicity, a function $F(\mathcal{E})$ is defined as 
\begin{equation}
F(\mathcal{E} ) = \frac{{{\mathcal{E}  - E} }}{{2\left(\mathcal{E}  - D\right)}},
\label{eq:3}
\end{equation}
For energies above $D$, $\lambda$ equals the function $F$ evaluated for $\mathcal{E}=E_C$ and 
 $\lambda=1-F(E_V)$ otherwise.
The space charge density of the Poisson equation is obtained by summing the electron or hole charge density with the background doping density.
If not stated otherwise, the Poisson equation is solved in ECA with dielectric constants of the respective material. 
Only for silicon UTB devices the thickness dependent dielectric constant of Ref.~[\onlinecite{Markov2015}] is used.
Note that the interpolation of the ECA model only applies to band-to-band tunneling situations. 
This can lead to inconsistencies when both, band-to-band tunneling and intra-band transport are relevant for the  device performance (e.g. p/n junctions in forward bias).

\begin{table}[h]
\caption{\label{tab:heuristics_table}The heuristic interpolation factor $\lambda(E)$ of the three heuristic ECA models in band-to-band tunneling devices. All given formulas and numbers are dimensionless. The interpolation factor $\lambda(E)$ is applied when the energy is above the valence band edge and below the conduction band edge. Below (above) the valence (conduction) band edge, $\lambda(E)$ is equal $1$ ($0$). If transport happens exclusively in valence or conduction bands (e.g. in MOSFETs), no interpolation factor is applied in ECA.
}
\centering
\begin{ruledtabular}
\begin{tabular}{cccc}
Label & $\alpha$ & E $\le$ D & E $ > $ D\\ \hline \\ [-0.8em] 
A & $ 0.2 $ & $ \lambda=1-F(E_{V}) $ & $ \lambda=F(E_{C}) $\\ \\ [-0.6em]
B & $ 0.5 $ & $ \lambda=1-F(E_{V}) $ & $ \lambda=F(E_{C}) $\\ \\ [-0.7em]
C & $ 0.5 $ & $1$ & $0$
\end{tabular}
\end{ruledtabular}
\end{table}

As a consistent alternative to the ECA model, this work extends the charge self-consistent model (termed as full-band approach or FBA) of Refs.~[\onlinecite{Andlauer2009}, \onlinecite{Areshkin2010}] to atomistic quantum transport of band-to-band tunneling devices within the NEGF formalism. 
For the sake of completeness, we repeat the model details here.
Every state solved within the quantum transport method is considered electronic and contributes, if occupied, a negative unit charge. 
This is irrespective of which band that state is in.
However, this model requires resolving the density contribution of all occupied states.

The standard NEGF treatment of density calculations requires to integrate the diagonal of the retarded Green's function $G^{R}$ over an energy interval that covers all occupied states\cite{Lake1997,4618725}.
The recursive Green's function implementation of NEMO5 is applied to solve for the diagonal of $G^{R}$.
Green's functions and self-energies are matrices in the position space indicated in bold font.
Most of the following equations involve the diagonal of the Green's functions only which is denoted in nonbold letters. 

\begin{figure}[t]
\centering
\includegraphics[width=0.4\textwidth,keepaspectratio]{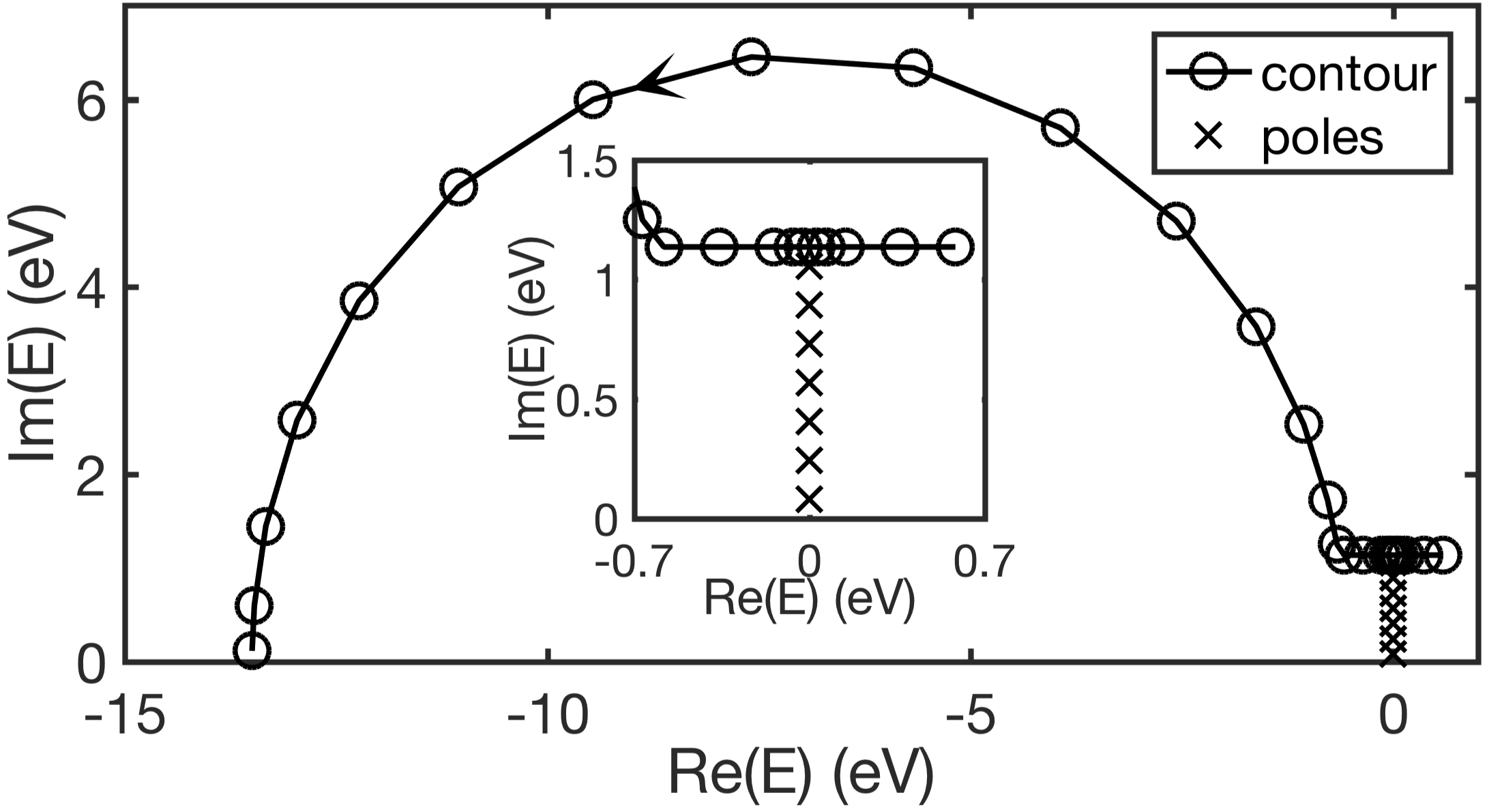}
\caption{\label{fig:contour} A typical integration contour used for $n_eq$ in Eq.~(\ref{eq:4}). The Fermi level of the left contact $\mu_{L}$ is set to $0$. Poles enclosed by the contour are marked by crosses and highlighted in the inset. The inset also illustrates the dense distribution of mesh points around the Fermi level that ensure a well resolved contact Fermi function. The arrow on the contour indicates the direction of the integral in the complex plane.}
\end{figure}

The total electron density $n_{FBA}$ is separated into an equilibrium $n_{eq}$ and a nonequilibrium part $n_{neq}$.
\begin{equation}
{n_{FBA}} = {n_{eq}} + {n_{neq}}.
\label{eq:4}
\end{equation}
The equilibrium electron density contribution is dependent on one contact Fermi function\cite{PAPIOR20178} (e.g. the left one) and is given by
\begin{equation}
{n_{eq}} = \sum\limits_k {\displaystyle\int\limits_{ - \infty }^\infty  {\frac{{ - {\mathop{\rm Im}\nolimits} \left[ {{G^R}(E,k)} \right]}}{\pi }} {f_L}(E,{\mu _L})dE},
\label{eq:5}
\end{equation}
where $\mu_L$ is the Fermi level of the left contact. 
Many atomistic models yield 10s of eV with hundreds of van Hove singularities of fully occupied valence bands\cite{PhysRevB.92.085301}, which are all considered within $n_{eq}$.
To avoid resolving all these states on a real energy mesh which poses immense numerical loads\cite{PhysRevLett.82.1225}, $n_{eq}$ is solved with the Residual theorem\cite{ZELLER1982993}
\begin{eqnarray}
{n_{eq}}=&&\sum\limits_k \left\{ \displaystyle\int\limits_{H + C} {\frac{{{\mathop{\rm Im}\nolimits} \left[ {{G^R}(E,k)} \right]}}{\pi }} {f_L}(E,{\mu _L})dE\right.\nonumber\\ 
&&\left.
+i2{k_B}T\sum\limits_{pole} {{G^R}({E_{pole}},k)}  \right\}. 
\label{eq:6}
\end{eqnarray}
The poles of the integrand originate from the Fermi function of the (left) contact.
These poles are located at $E_{pole}=\mu_L+ik_BT\pi(2m+1)$ with $Res(E_{pole})=-k_BT$, ${m\in\mathbb{N}}$. 
A typical integration contour \cite{Areshkin2010,Brandbyge2002} is shown in Fig.~\ref{fig:contour}. 
The integration contour consists of a semicircular part (C) whose lower bound is set about $1 \, eV$ below the lowest eigenvalue of the system.
The horizontal part of the contour (H) is parallel to the real energy axis.
The maximum real part of H is exceeding $\mu_L$ by $25 \, k_BT$ to include the complete tail of the contact Fermi function in the density calculation. 
The small contour portion that closes the integration contour beyond the horizontal section H does not have a net contribution to Eq.~(\ref{eq:5}). 
When the imaginary part of the integration contour is large enough (such as indicated in Fig.~\ref{fig:contour}), numerical solutions of Eq.~(\ref{eq:6}) converge with few tens of contour points. 

The integral of the non-equilibrium electron density must be performed along the real energy axis since the integrand is not analytic in the entire complex plane. 
\begin{eqnarray}
&& n_{neq} = \nonumber \\
&& \sum\limits_k \displaystyle\int\limits_{ - \infty }^\infty  diag\left\{\bm{{G^R}(E,k)}\frac{{{\mathop{\rm Im}\nolimits} \left[ {\bm{\Sigma ^R(E,k)}} \right]}}{\pi } \bm{G^R(E,k)^\dag} \right\} \nonumber \\
&&\times
\left[ {{f_L}(E,{\mu _L}) - {f_R}(E,{\mu _R})} \right]dE.
\label{eq:7}
\end{eqnarray}
Equation~(\ref{eq:7}) involves a matrix product of Green's functions and a self-energy indicated by bold letters.
The integration window is restricted by the two contact Fermi functions ($\mu_R$ being the right contact's Fermi level) and is approximately the same energy window considered in the ECA. 
Compared to the ECA, the extra computational load in FBA is given by a few tens of energy points for the integral of the equilibrium electron density in Eq.~(\ref{eq:6}). 
This is a negligible addition given the  hundreds or thousands of energy points typically needed to resolved the non-equilibrium density contribution $n_{neq}$.

In charge self-consistent FBA calculations, the Poisson equation requires a positive background charge $n_{core}$ to allow the presence of electrons in the devices. 
This background charge is assumed to completely compensate the electronic charge density of the respective undoped device in equilibrium.
The total space charge $\rho_{FBA}$ is given by the sum of $n_{eq}$, $n_{neq}$, $n_{core}$ and the doping density $n_{doping}$.
\begin{equation}
\rho_{FBA} = - q{n_{FBA}} + q{n _{core}} + q{n_{doping}},
\label{eq:8}
\end{equation}

The dielectric constant of vacuum is used in the Poisson equation in FBA since the screening of all valence band electrons is explicitly included in the calculation\cite{Niquet2011}. 
As will be shown in Sec. III, this feature of FBA is important since it makes the FBA model independent of the material and device specific dielectric screening. 
The charge self-consistent loop with an approximate Jacobian that takes into account the response of both the equilibrium and the non-equilibrium electron density to potential changes turns out to converge typically within 15 iterations.

\section{\label{sec:level1}RESULTS}
\subsection{Convergence behavior of $n_{eq}$ calculation}
\begin{figure}[H]
\centering
\includegraphics[width=0.49\textwidth,keepaspectratio]{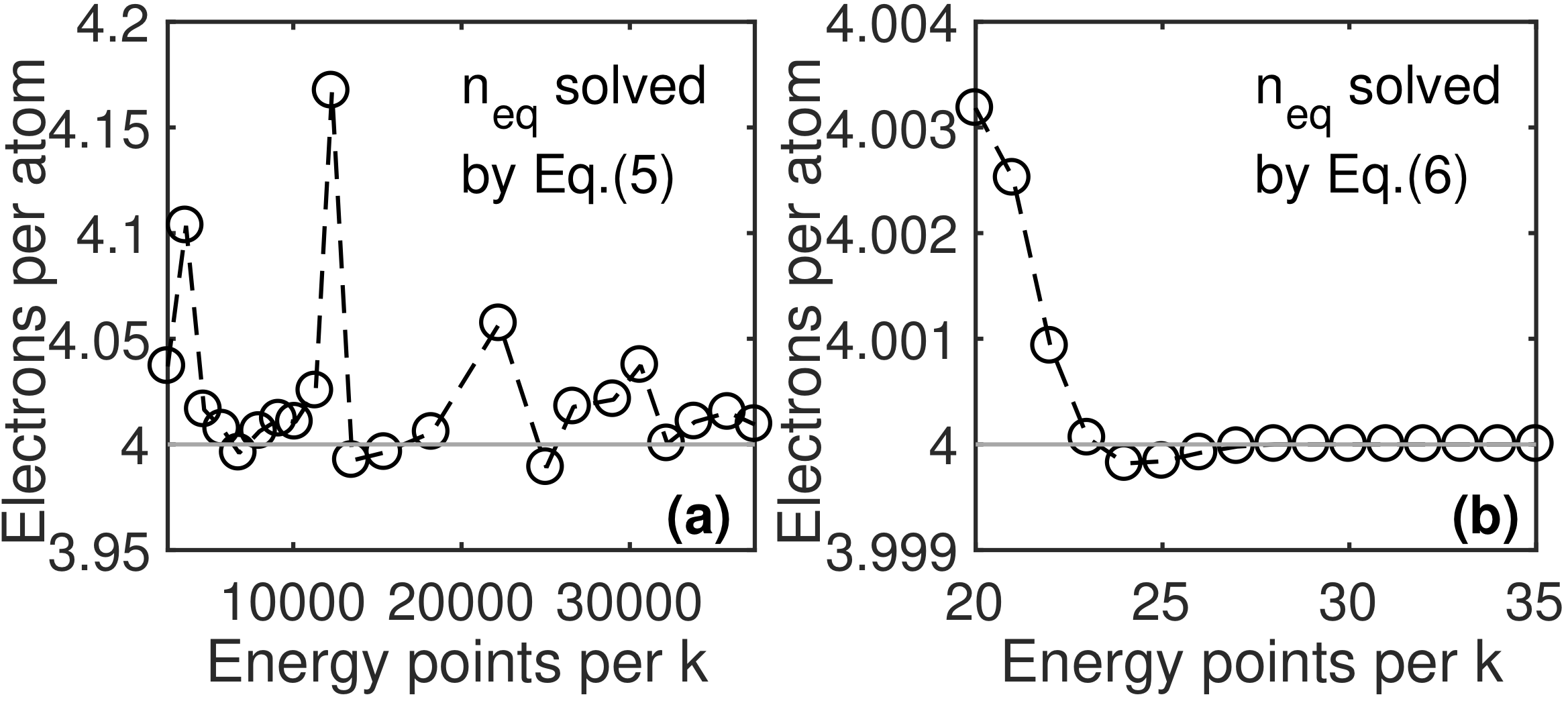}
\caption{\label{fig:convergence_complexand_real} The atom resolved electron density $n_{eq}$ vs number of energy points per momentum point for the real energy integral in  Eq.~(\ref{eq:5}) (a) and for the complex contour integral of Eq.~(\ref{eq:6}) (b). The converged number of 4 valence Si electrons is indicated with a solid line to guide the eye. The $n_{eq}$ converges with only 30 energy points in Eq.~(\ref{eq:6})  while no convergence is observed when solved along the real energy axis even with an immense number of energy points.}
\end{figure}

The numerical convergence of solving the equilibrium electron density $n_{eq}$ with Eqs.~(\ref{eq:5}) and (\ref{eq:6}) with varying number of energy points per momentum point is compared for homogeneous 3D silicon in Figs.~\ref{fig:convergence_complexand_real} (a) and (b). 
In both cases, the electronic Brillouin zone is resolved with 225 momentum points. 
The equilibrium density of Eq.~(\ref{eq:6}) converges with only a few complex energy points (contour points and poles) per momentum point. 
In contrast, many and hard to resolve van Hove singularities on the real-energy axis prevent Eq.~(\ref{eq:5}) to fully converge even with an immense number of energy points.
\subsection{Transferability of the background charge}
\begin{figure}[H]
\centering
\includegraphics[width=0.3\textwidth,keepaspectratio]{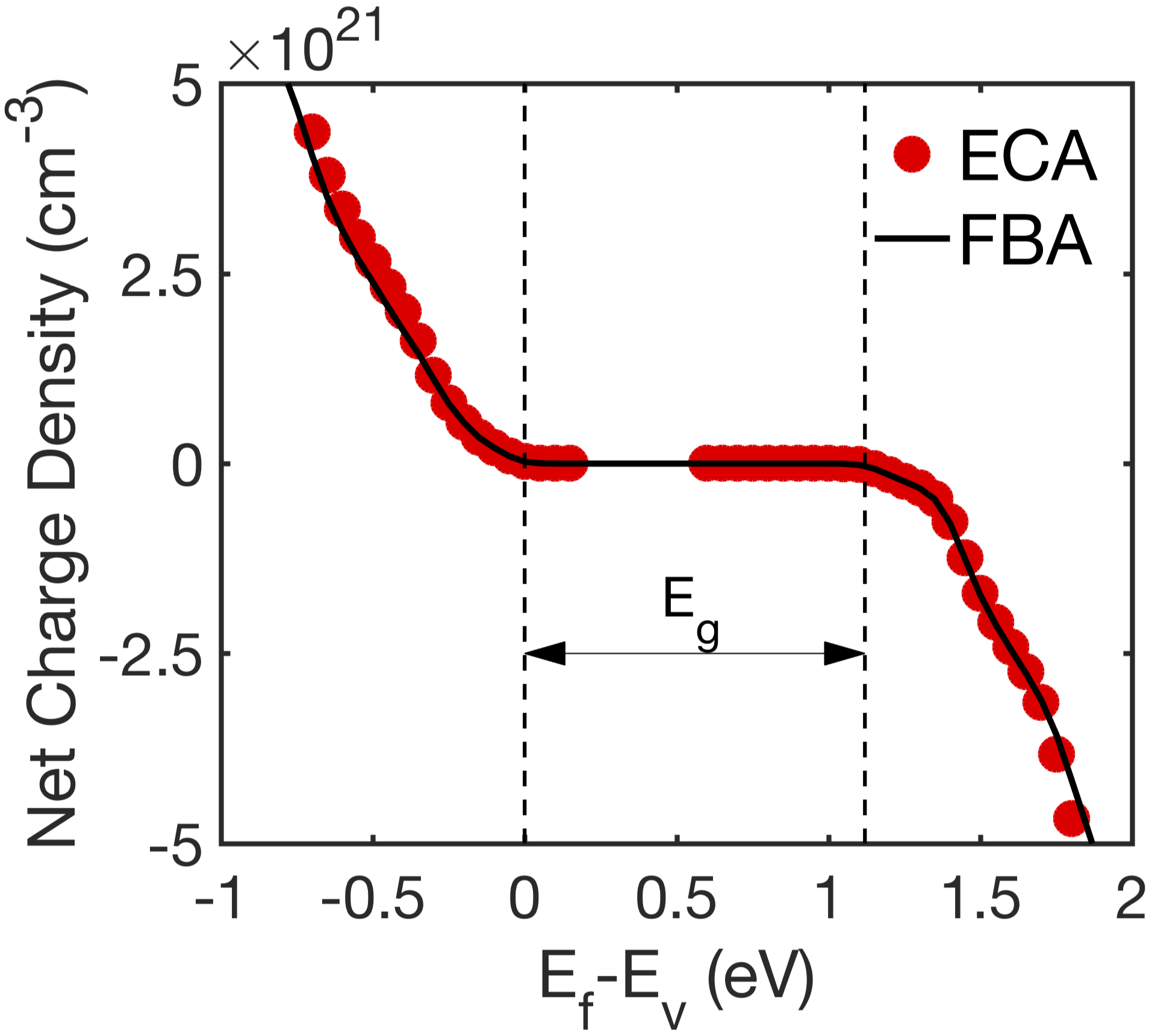}
\caption{\label{fig:net_charge_comparison} Variation of charge density with effective Fermi level  calculated with both approaches. The two dashed lines indicate the valence (at 0) and conduction band edges. Deviation between ECA and FBA increases when the Fermi level is deep in the valence band or the conduction band. Then an increasing number of van Hove singularities plague the convergence of ECA.}
\end{figure}

The background charge $qn_{core}$ of the FBA model is determined for the intrinsic device. 
To verify the transferability of $n_{core}$ to finite doping situations, the dependence of charge density on the effective Fermi level in the FBA model ($-n_{FBA}+n_{core}$) is compared in three dimensional silicon in equilibrium (see Fig.~\ref{fig:net_charge_comparison}) with predictions of the ECA model ($-n_{ECA}$). 
Screening in homogeneous silicon in equilibrium is well described with the material dielectric constant.
Accordingly, both models agree excellently and the background charge of the FBA model applies over a large range of density variations.
Deviation increases when the Fermi level is deep in the valence band or the conduction band, i.e. when the ECA model faces a lot of van Hove singularities and a converged solution of the real energy integral in Eq.~(\ref{eq:5}) becomes numerically challenging.
\subsection{Transfer characteristics of silicon MOSFETs}
Transfer characteristics resulting from FBA and ECA charge self-consistent calculations of a p-type and a n-type silicon ultra-thin body double-gate MOSFET are shown in Fig.~\ref{fig:nin_pip_IV_and_potential}.
Both MOSFET devices have a structure with body thickness, channel and source/drain lengths of $1.6 \, nm$, $10.8 \, nm$ and $11.4 \, nm$, respectively. 
A doping concentration of $1 \times 10^{20} \, cm^{-3}$ is assumed in the source and drain regions of both transistor types. 
A drain-to-source voltage $V_{DS}$ of $0.4 \, V$ is applied. 
The equivalent oxide thickness (EOT) of top and bottom oxides is $1 \, nm$. 
It is reported both experimentally\cite{PhysRevB.77.115355} and theoretically\cite{Markov2015} that the dielectric constant of silicon ultra-thin films reduces with the film thickness. 
A dielectric constant of $9.9$ is used in the ECA simulations following Ref.~[\onlinecite{Markov2015}]. 
In FBA calculations, only the vacuum dielectric constant enters the Poisson equation.
\begin{figure}[t]
\centering
\includegraphics[width=0.33\textwidth,keepaspectratio]{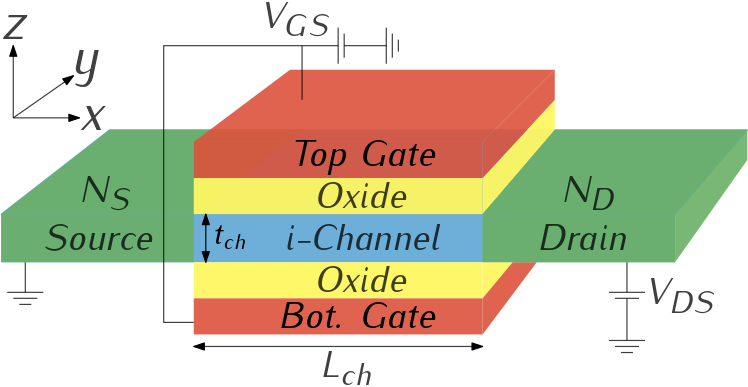}
\caption{\label{fig:device_structure} Schematic of the ultra-thin body double-gate transistors used in all transfer characteristic predictions of this work. The gate bias is controlled by $V_{GS}$. Electron transport occurs along $x$ direction when a non-zero $V_{DS}$ is applied. The channel of the device is confined along z direction and $t_{ch}$ is the channel thickness. Periodic boundary condition is assumed along y direction. $L_{ch}$ is the channel (gate) length. $N_{S}$ and $N_{D}$ are doping concentrations in source and drain regions, respectively.}
\end{figure}

Results of both models, for the transfer characteristics and band profiles agree very well. 
This is particularly true compared to TFET situations (see following subsections) that define the common ECA/FBA difference scale. 
The good agreement is expected since electronlike and holelike states are clearly separated in these devices. 
The maximum relative difference of the drain current in the two models is below $15\%$ and the maximum potential difference is below $k_BT$. 
Note that the dielectric constant of the ECA model could serve as a fitting degree of freedom to match the FBA results. 
This finding emphasizes the strength of the FBA model to explicitly handle the electrostatic response of the deep lying valence electrons - with a marginal increase in computational cost.
\begin{figure}[H]
\centering
\includegraphics[width=0.49\textwidth,keepaspectratio]{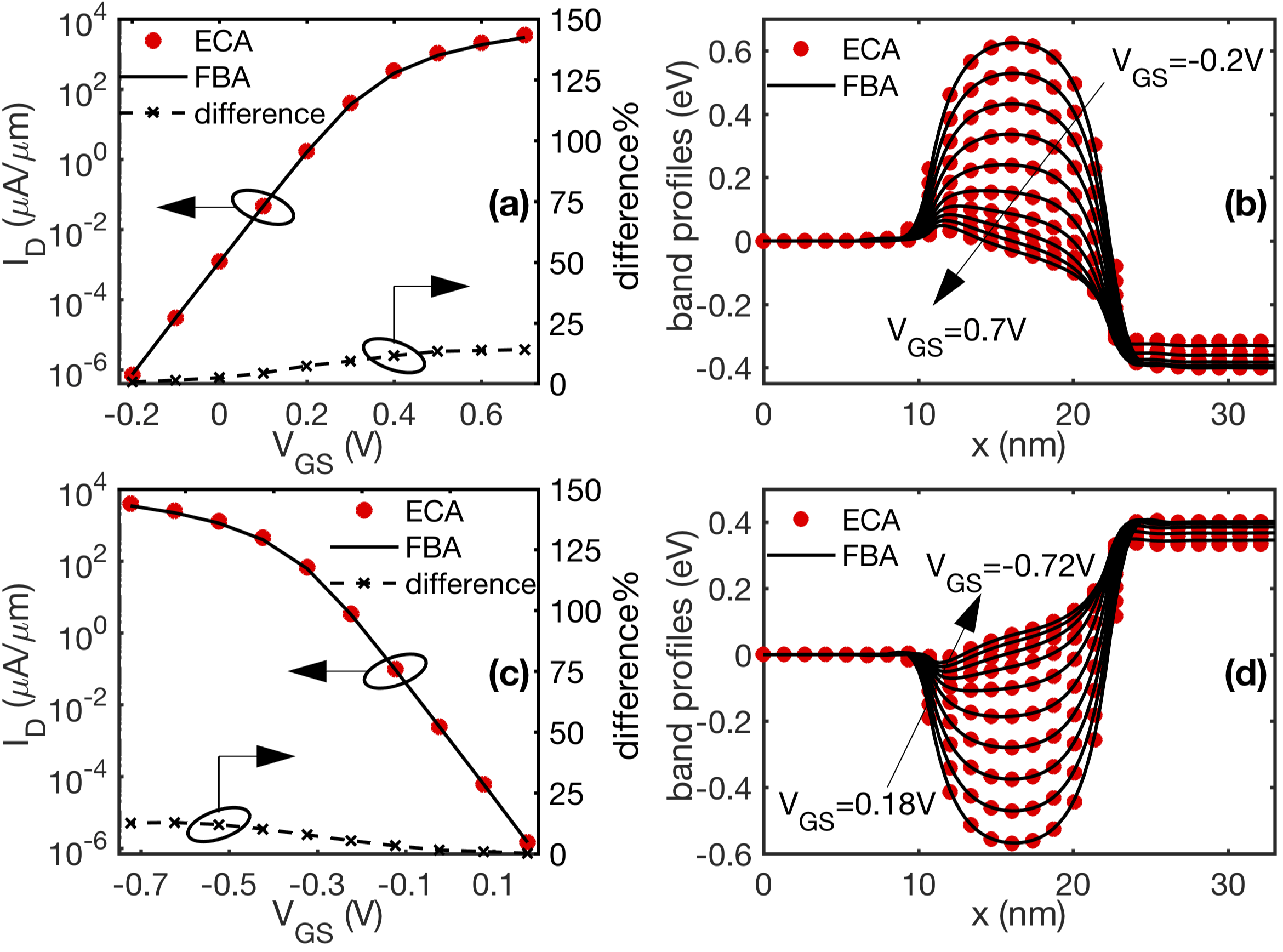}
\caption{\label{fig:nin_pip_IV_and_potential} Transfer characteristics $I_D-V_{GS}$ of silicon ultra-thin body double-gate nMOSFET (a) and pMOSFET (c) of Fig.~\ref{fig:device_structure} predicted with ECA and FBA. The percentage difference is plotted in dashed lines with cross markers. (b) and (d) Band profiles of ECA and FBA corresponding to $V_{GS}$ nodes in (a) and (c), respectively.}
\end{figure}
\subsection{Silicon TFET}
\begin{figure}[H]
\centering
\includegraphics[width=0.43\textwidth,keepaspectratio]{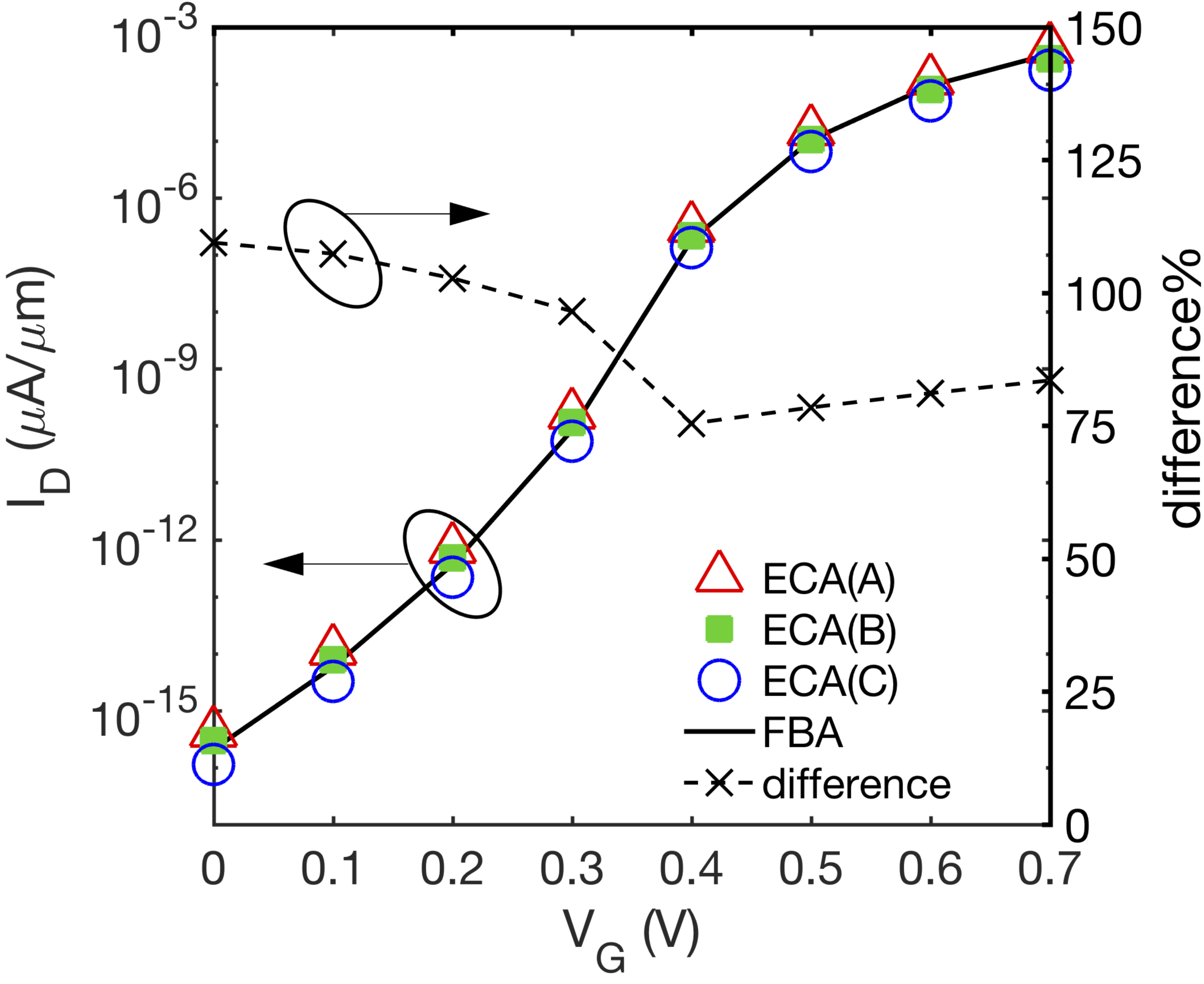}
\caption{\label{fig:Id_Vg_Si_TFET} Transfer characteristics $I_D-V_{GS}$ at $V_{DS} = 0.3 \, V$ of a silicon ultra-thin body double-gate TFET. Results of FBA and ECA with three different heuristic models are shown. The maximum deviation of the three ECA results relative to their average is plotted in dashed line with cross markers.}
\end{figure}

To compare the predictions of ECA and FBA for BTBT devices, the device structure of Fig.~\ref{fig:device_structure} is considered with the source (drain) doping being p-type (n-type). 
The source and drain doping concentrations are $5 \times 10^{19} \, cm^{-3}$ and $2 \times 10^{19} \, cm^{-3}$, respectively. 
The body thickness, channel and source/drain lengths are $1.6 \, nm$, $10.8 \, nm$ and $11.4 \, nm$, respectively. 
A dielectric constant of $9.9$ is used in the ECA model following Ref.~[\onlinecite{Markov2015}]. 
The ECA utilizes heuristic models to distinguish electronlike and holelike charge contributions in BTBT devices. 
Three commonly used heuristic models (summarized in Table.~\ref{tab:heuristics_table})) are applied in ECA and results are presented and compared to the FBA result (see Fig.~\ref{fig:Id_Vg_Si_TFET}).

\begin{figure}[t]
\centering
\includegraphics[width=0.4\textwidth,keepaspectratio]{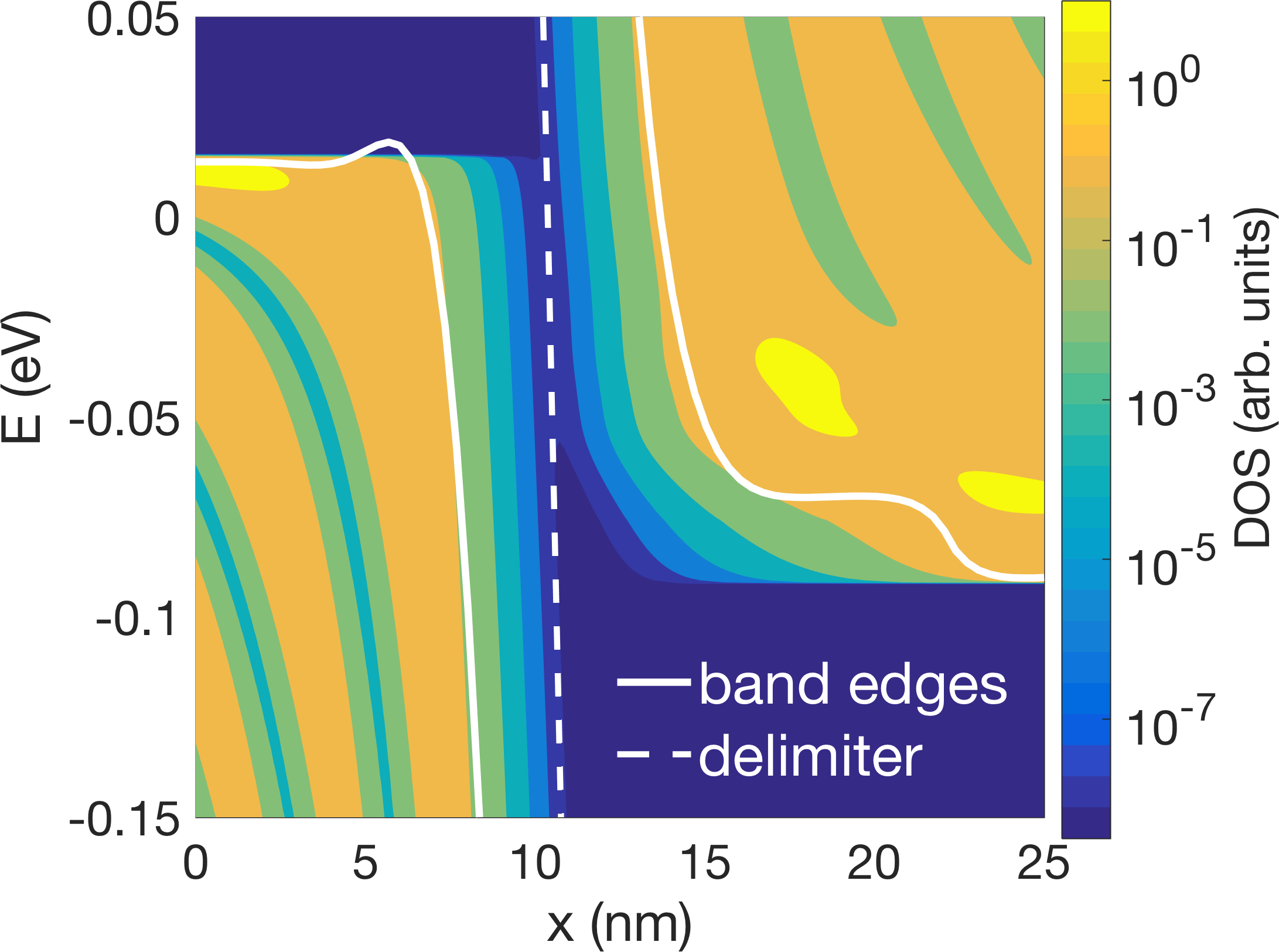}
\caption{\label{fig:Si_TFET_band_edges_dos_and_delimiter} Contour plot of the energy and position resolved density of states of the silicon TFET simulated in Fig.~\ref{fig:Id_Vg_Si_TFET} at $k=\Gamma$. The conduction and valence band edges are represented in white solid lines. The white dashed line depicts a hole/electron delimiter for $\alpha = 0.5$ in the band gap which is used to distinguish electron and hole states in ECA. }
\end{figure}

The difference in the performance predictions of ECA and FBA can be understood from Fig.~\ref{fig:Si_TFET_band_edges_dos_and_delimiter}. 
The electron density in the bandgap (at around $10 \, nm$) is considered as electrons in FBA, whereas in ECA, a charge prefactor is assigned to it. 
This prefactor depends on the position of the hole/electron delimiter and the considered interpolation scheme (see Table.~\ref{tab:heuristics_table}).
A snapshot of the delimiter for $\alpha = 0.5$ is illustrated in Fig.~\ref{fig:Si_TFET_band_edges_dos_and_delimiter}. 
Consequently, that prefactor differs in the three applied heuristic models.
The different prefactors in turn impact the interpretation of ECA charge and the electrostatic potential around the tunnel junction.
Thus, the TFET transfer characteristic prediction is sensitive to the chosen ECA delimiter model. 
This is indicated by the dashed line with cross markers in Fig.~\ref{fig:Id_Vg_Si_TFET} which shows the maximum deviation among all ECA models relative to their average.

\subsection{p-GaSb/n-InAs HJTFET}

\begin{figure}[H]
\centering
\includegraphics[width=0.43\textwidth,keepaspectratio]{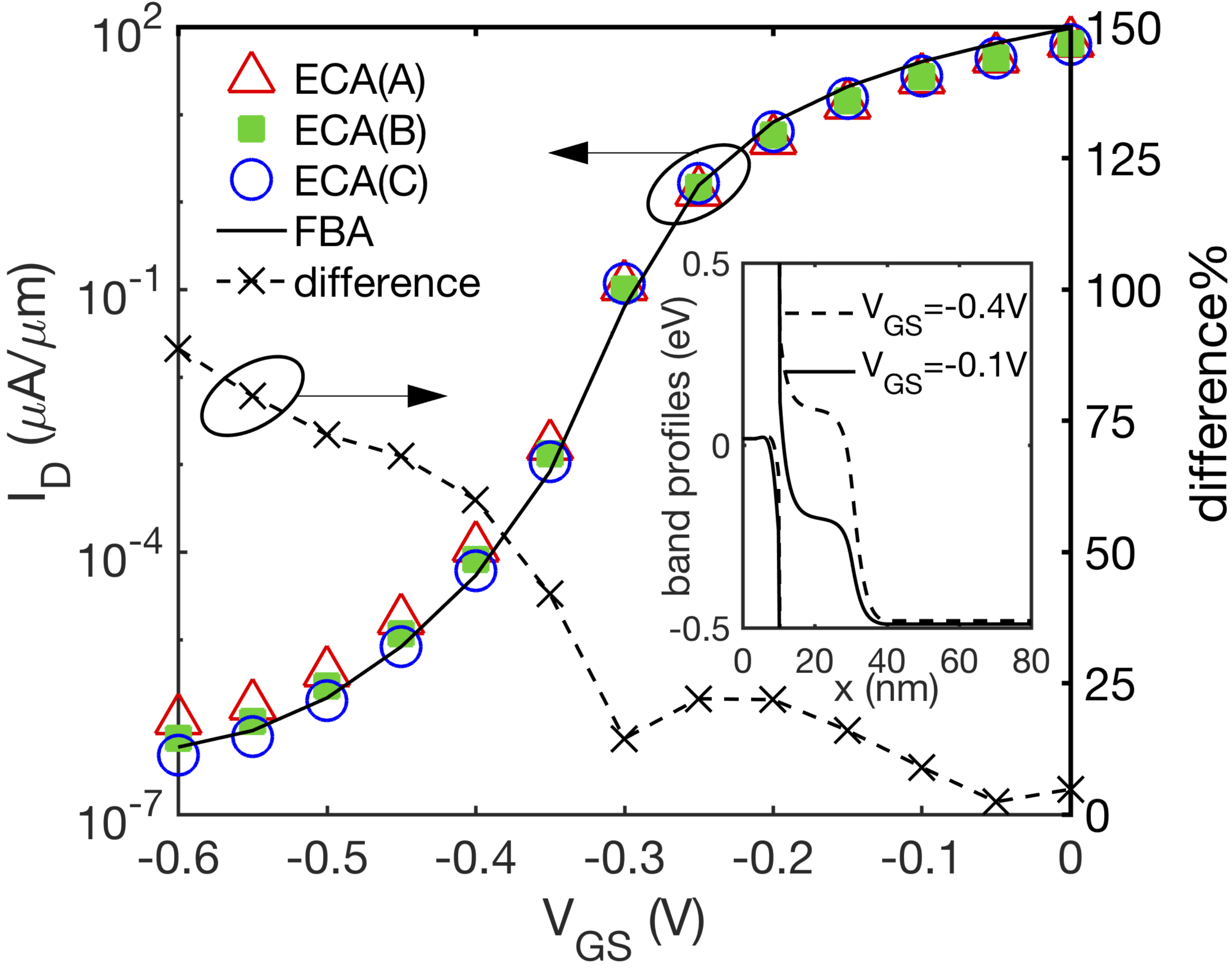}
\caption{\label{fig:Id_Vg_comparison_EC_and_EH_with_partial_different_heuristics_GaSb_InA} 
Transfer characteristics $I_D-V_{GS}$ at $V_{DS} = 0.3 \, V$ of a double-gate p-GaSb/n-InAs heterojunction TFET with channel thickness $t_{ch}=3 \, nm$ predicted with FBA and ECA with three different heuristic models. The maximum deviation of the three ECA results relative to their average is plotted in dashed line with cross markers. The inset shows band profiles in the middle of the TFET for two gate voltages ($V_{GS} = -0.4 \, V$ and $V_{GS} = -0.1 \, V$).}
\end{figure}

For completeness, the predictions of FBA and the three different ECA models are compared on a direct bandgap heterojunction TFET as well.
The transfer characteristics of a double-gate p-GaSb/n-InAs heterojunction TFET with channel thickness $t_{ch}=3 \, nm$ is shown in Fig.~\ref{fig:Id_Vg_comparison_EC_and_EH_with_partial_different_heuristics_GaSb_InA}.
The channel length of the considered TFET is  $L_{ch}=20 \, nm$. 
The doping density in p-GaSb source and n-InAs drain regions are $5 \times 10^{19} \, cm^{-3}$ and $2 \times 10^{19} \, cm^{-3}$, respectively. 
Short of dielectric constant assessments for GaSb and InAs UTBs, the respective material constants are used within ECA. 
A drain-to-source voltage $V_{DS}$ of $0.3 \, V$ is applied. 
The equivalent oxide thickness (EOT) of top and bottom oxides is set to $0.47 \, nm$. 

The relative variation of the ECA predicted drain current densities decreases with increasing gate voltage increases. 
This can be understood with the comparison of band profiles for different gate voltages shown in the inset of Fig.~\ref{fig:Id_Vg_comparison_EC_and_EH_with_partial_different_heuristics_GaSb_InA}. 
The effective tunneling distance, i.e. the spatial area in which ECA-specific models of Table~\ref{tab:heuristics_table} are applied, shrinks with increasing gate voltage.
Then, differences in the ECA models become less relevant as well.

\section{\label{sec:level1}SUMMARY}
In this work, the FBA of Andlauer and Vogl is adapted to the framework of nonequilibrium Green's function based charge self-consistent quantum transport in atomically resolved nanotransistors. 
This model is compared to state of art ECA that heuristically interpret band-to-band tunneling states as partly electronlike and holelike. 
Significant ambiguity of the heuristic ECA is exemplified on various tunneling field effect transistors. 
The FBA lifts this ambiguity by explicitly including all valence band states and consistently interpreting them as electronlike only.
For conventional MOSFETs that lack significant band to band tunneling, the ECA agrees with the FBA only if an appropriate screening constant for the valence electrons is applied.
This constant is known to deviate from the respective material value. 
Since the FBA considers all valence electrons explicitly, no extra input of screening constants is required.
In conclusion, the FBA provides a much wider application space than the conventional ECA - with a marginal increase in computational cost.
\begin{acknowledgments}
This work is supported by NSF EFRI-1433510 and in part by funding from the Semiconductor Research Corporation (GRC 2653.001). This work is also supported in part by Intel Parallel Computing Center at Purdue. We also acknowledge the Rosen Center at Purdue University. This work is part of the Blue Waters sustained-petascale computing project, which is supported by the National Science Foundation (award number ACI 1238993) and the state of Illinois. Blue Waters is a joint effort of the University of Illinois at Urbana-Champaign and its National Center for Supercomputing Applications. Finally, we would like to thank Yann-Michel Niquet for helpful discussions.
\end{acknowledgments}
\appendix
\section{Incompatibility of ECA with NEGF}
The following derivations show the ECA model is not applicable to the NEGF formalism since it violates fundamental NEGF equations. 
The retarded, advanced, lesser and greater Green functions are linearly dependent\cite{haug2007quantum}
\begin{equation}
\bm{{G^R}(E,k)} - \bm{{G^A}(E,k)} = \bm{{G^ > }(E,k)} - \bm{{G^ < }(E,k)}.
\label{eq:A1}
\end{equation}
This equation holds for the full Green's function matrices, but for simplicity, the following derivations focus on the diagonals only.
The real part of the spectral function $A$ equals the density of states. 
It is defined as\cite{Datta2000}
\begin{equation}
\begin{array}{cl}
A(E,k) &\equiv i\left[ {{G^R}(E,k) - {G^A}(E,k)} \right] \\
&= i\left[ {{G^ > }(E,k) - {G^ < }(E,k)} \right].
\end{array}
\label{eq:A2}
\end{equation}
In equilibrium, the fluctuation-dissipation theorem gives\cite{haug2007quantum} 
\begin{eqnarray}
{G^ > }(E,k) \: &&=  - i\left[ {1 - f(E)} \right]A(E,k), \label{eq:A3} \\
{G^ < }(E,k) \: &&= if(E)A(E,k). \label{eq:A4}
\end{eqnarray}
Since the lesser (greater) Green's function is related to the electron (hole) density\cite{Lake1997}, the extension of the ECA interpolation of Eq.~(\ref{eq:1}) to the NEGF framework would require Eqs.~(\ref{eq:A3}) and (\ref{eq:A4}) to read
\begin{eqnarray}
{G^ > }(E,k) \: &&=  - i\left[ {1 - f(E)} \right]A(E,k)\lambda(E), \label{eq:A5} \\
{G^ < }(E,k) \: &&= if(E)A(E,k)\left[ {1 - \lambda(E)} \right], \label{eq:A6}
\end{eqnarray}
When Eqs.~(A5) and (A6) are inserted into Eq.~(A2), the spectral function has to fulfill
\begin{equation}
\begin{array}{cl}
A(E,k) &= \left[ {1 - f(E)} \right]A(E,k)\lambda (E) \\
&\quad + f(E)\left[ {1 - \lambda (E)} \right]A(E,k).
\end{array}
\label{eq:A7}
\end{equation}
This requires
\begin{equation}
\lambda(E) - 2\lambda(E)f(E) + f(E) = 1,
\label{eq:A8}
\end{equation}
which cannot hold true in general.

Note that altering the Green's function matrices beyond their diagonals shows another shortcoming of the ECA model.
The two propagation coordinates of the Green's functions can yield very different interpolation values.
It is unclear which value the nonlocal Green's function elements should be altered with.
\bibliography{aipmain_reference}

\end{document}